\documentclass[12pt]{article}
\usepackage[usenames]{color}
\usepackage{amsmath}
\usepackage{amsfonts}
\usepackage{amssymb}

\newcommand{\be}{\begin{equation}}
\newcommand{\ee}{\end{equation}}
\newcommand{\bea}{\begin{eqnarray}}
\newcommand{\eea}{\end{eqnarray}}
\newcommand{\beaa}{\begin{eqnarray*}}
\newcommand{\eeaa}{\end{eqnarray*}}

\newcommand{\nn}{\nonumber}

\def\p{\partial}

\def\l{\lambda}
\begin{document}
\title{Grassmannians Gr(N-1,\,N+1), closed differential N-1 forms and 
N-dimensional integrable systems}
\author{
L.V. Bogdanov\thanks
{L.D. Landau ITP RAS,
Moscow, Russia}
~and B.G. Konopelchenko
\thanks{Department of Mathematics and Physics "Ennio De Giorgi", University of
Salento and INFN, Lecce, Italy}
}
\date{}
\maketitle
\begin{center}
\textit{To the memory of S.V. Manakov}
\end{center}
\begin{abstract}
Integrable flows on the Grassmannians Gr(N-1,\,N+1) are defined by the
requirement of closedness of the differential N-1 forms $\Omega _{N-1}$ of
rank N-1 naturally associated with  Gr(N-1,\,N+1). Gauge-invariant parts of
these flows, given by the systems of the N-1 quasi-linear differential
equations, describe coisotropic deformations of (N-1)-dimensional linear
subspaces. For the class of solutions which are Laurent polynomials in one
variable these systems coincide with  N-dimensional  integrable systems
such as Liouville equation (N=2), dispersionless Kadomtsev-Petviashvili
equation (N=3), dispersionless Toda equation (N=3), Plebanski second
heavenly equation (N=4) and others. Gauge invariant part of the forms 
$\Omega _{N-1}$ provides us with the compact form of the corresponding
hierarchies. Dual quasi-linear systems associated with the projectively
dual Grassmannians Gr(2,\,N+1) are defined via the requirement of  the
closedness of the dual forms $\Omega _{N-1}^{\star}$.  It is shown that at
N=3 the self-dual quasi-linear system, which is associated with the harmonic
(closed and co-closed) form $\Omega _{2}$,  coincides with the Maxwell
equations for orthogonal electric and magnetic fields. 
\end{abstract}
\section{ Introduction}
Multidimensional integrable systems which are equivalent to the
commutativity of multidimensional vector fields
\begin{equation}
D_{\alpha }=\sum_{k=0}^{N}a_{\alpha k}(x;\lambda)\frac{\partial }{\partial
x_{k}},\qquad \alpha =1,2
\end{equation}%
where $a_{\alpha k}(x;\lambda )$ are rational functions of the spectral
parameter $\lambda $ have been invented by Zakharov and Shabat in the
seminal paper \cite{ZS}. \ Concrete examples of such systems and an
extension of this scheme to the case when operators $D_{\alpha }$ contain
also the derivatives with respect to spectral parameter $\lambda $ 
($\frac{\partial}{\partial \lambda}=\frac{\partial }{\partial x_{0}}$) have been
considered in the papers \cite{BZ78,BZM87}, \cite{Pleb75}, 
\cite{Takasaki89,Takasaki89a},
\cite{KodGib90,Zakh94,Krich94}, \cite{MS06,MS07}, \cite{TT91,TT95}, 
\cite{Dun02}. 
This class of multidimensional integrable
systems includes second heavenly equation \cite{Pleb75}, 
hyper-K\"ahler hierarchy \cite{Takasaki89,Takasaki89a},
dispersionless Kadomtsev-Petviashvili (dKP) equation 
\cite{KodGib90,Zakh94,Krich94},
Manakov-Santini
system \cite{MS06,MS07},
Boyer-Finley (dispersionless Toda) equation \cite{TT91,TT95}, 
Dunaski equation \cite{Dun02} and their generalisations.  In recent years various
properties of such integrable systems (wave breaking, reductions, etc.)
have been intensively studied \cite{MS08,MS09,MS11,LVB10MS,LVB11,LVB12,{BDM07},
{LVB09},{LVB10Toda}}. 
In particular, in the paper \cite{LVB10MS} it
was observed that in addition to the equations for the variables $a_{\alpha
k}(x)$ the equations for the Jacobian $J$ of the solutions $\Psi$ of the
systems $D_{\alpha }\Psi =0$, $\alpha =1,2$ play an important role in the whole
method. For instance, it allows to get a compact form of the corresponding
hierarchies.

In the present paper we show that all these integrable systems arise
naturally in the families of the Grassmannians Gr(N-1,N+1) parameterized by
N+1 variables $x_{0},\,x_{1},\,\dots,\,x_{N}$.  Namely, they are equivalent to
the closedness conditions for differential N-1 forms
\begin{equation}
\Omega _{N-1}=\omega _{0}\wedge \dots \wedge \omega _{N-2}
\end{equation}
where 1-forms $\omega _{\beta }=\sum_{i=1}^{N+1}p_{i}^{\beta }dx_{i}$, $\beta
=0,\dots,N-2$ are built of N-1 vectors $p_{\beta }$ defining (N-1)-dimensional
linear subspaces. \ In terms of Pl\"ucker coordinates 
$\pi_{i_{0}\,i_{1}\,\dots\,i_{N-2}}$, 
$i_{k}=0,1,\dots,N$ one has the system of linear
equations
\begin{equation}
\left[ \frac{\partial \pi _{i_{0}\,i_{1}\,\dots i_{N-2}}}{\partial x_{i_{N-1}}}%
\right] =0.
\label{closedform}
\end{equation}%
This system is equivalent to the system of N+1 quasilinear equations
\begin{equation}
\frac{\partial J}{\partial x_{N-1}}+\sum_{l=0}^{N-2}\frac{\partial (Ja_{1m})%
}{\partial x_{m}}=0,\quad \frac{\partial J}{\partial x_{N}}+\sum_{l=0}^{N-2}%
\frac{\partial (Ja_{2m})}{\partial x_{m}}=0,
\label{subsystemlin}
\end{equation}
\begin{equation}
\frac{\partial a_{1k}}{\partial x_{N}}-\frac{\partial a_{2k}}{\partial
x_{N-1}}+\sum_{l=0}^{N-2}\left( a_{2l}\frac{\partial a_{1k}}{\partial x_{l}}%
-a_{1l}\frac{\partial a_{2k}}{\partial x_{l}}\right) =0,\qquad k=0,1,...,N-2.
\label{subsystem}
\end{equation}%
for $J=\pi _{0,1,...,N-2}$ and 2(N-1) independent affine coordinates
\[
a_{1k}=(-1)^{k}J^{-1}\pi _{0\,\dots\,k-1\,k+1\,\dots\,N-1},\quad
a_{2k}=(-1)^{k}J^{-1}\pi
_{0\,\dots\,k-1\,k+1\,\dots\,N}, 
\]%
where $k=0,\dots,N-2$
The subsystem (\ref{subsystem}) 
coincides with the system describing the coisotropic
deformations of the N-1 dimensional linear space defined by the equations
\begin{equation}
p_{N-1}+\sum_{k=0}^{N-2}a_{1k}p_{k}=0,\qquad
p_{N}+\sum_{k=0}^{N-2}a_{2k}p_{k}=0.
\end{equation}%
The system (\ref{closedform}) 
provides us with the linearization of nonlinear
equation (\ref{subsystem}) in terms of the variables 
$J$, $Ja_{1m}$, $Ja_{2m}$, where $J$ is a
solution of equations (\ref{subsystemlin}).

For the class of solutions polynomial in the variable $x_{0}(=\lambda )$ the
system (\ref{subsystem}) becomes 
the system of the integrable N-dimensional equations of
the type considered in the papers \cite{Pleb75}, \cite{Takasaki89,Takasaki89a},
\cite{KodGib90,Zakh94,Krich94}, \cite{MS06,MS07}, \cite{TT91,TT95}, 
\cite{Dun02}, \cite{{BDM07},{LVB09}}. 
Hierarchies of systems arising in
such a way can be represented in the compact form
\[
\left( J^{-1}\Omega _{N-1}\right) _{-}=\left( J^{-1}d\Psi _{0}\wedge d\Psi
_{1}\wedge \cdot \cdot \cdot \wedge d\Psi _{N-2}\right) _{-}=0 
\]%
where $\Psi _{0},\Psi _{1},\dots,\Psi _{N-2}$ are Darboux type cooordinates
for the closed N-1 form $\Omega _{N-1}$ which are the Laurent series in $%
x_{0}=\lambda $ and $(\cdots)_{-}$ denotes the projection on the negative part of
Laurent series. Simple generating formulae for the variables $\Psi _{k}$ are
also presented in \cite{LVB09}.

Duality between the Grassmannians \ Gr(N-1,N+1) and Gr(2,N+1) \ suggests to
consider the system \ dual to the system (\ref{subsystemlin},\ref{subsystem}).  
It is equivalent to the
closedness condition for the 2-form \ $\Omega _{2}^{\ast }=\star \Omega
_{N-1} $ where $\star $ denotes the Hodge star (or duality) operation or to
the co-closedness of the form $\Omega _{N-1}$ and it is of the form
\begin{equation}
\frac{\partial J^{\ast }}{\partial x_{k}}+\sum_{m=0}^{1}\frac{\partial
(J^{\ast }a_{km}^{\ast })}{\partial x_{m}}=0,\quad k=2,...,N,
\end{equation}
\begin{equation}
\frac{\partial a_{jl}^{\ast }}{\partial x_{k}}-\frac{\partial a_{kl}^{\ast }%
}{\partial x_{j}}+\sum_{m=0}^{1}\left( a_{km}^{\ast }\frac{\partial
a_{jl}^{\ast }}{\partial x_{m}}-a_{jm}^{\ast }\frac{\partial a_{kl}^{\ast }}{%
\partial x_{m}}\right) =0,\quad l=0,1;\quad j,k=2,...,N
\label{subsystemdual}
\end{equation}%
where $J^{\ast }=\pi _{01}^{\ast }$, $a_{j0}^{\ast }=J^{\ast -1}\pi
_{1j}^{\ast },a_{j1}^{\ast }=-J^{\ast -1}\pi _{0j}^{\ast }$ and $\pi
_{kj}^{\ast }$ are Pl\"ucker coordinates dual to $\pi
_{i_{0},i_{1,}...i_{N-2}} $ . \ At N=3 the system (\ref{subsystemdual}) 
is quite similar to
the original system (\ref{subsystem}). 
For N$\geq 4$ the system (\ref{subsystemdual}) 
is the semi-decoupled
system of $\frac{(N-2)(N-1)}{2}$ four-dimensional subsystems.

\bigskip Self-dual \ systems are associated with the harmonic forms $\Omega
_{N-1}$, i.e. the forms which are closed and co-closed. Self-dual system in
Gr(N-1,N+1) is the systems of $(N-1)^{2}$ equations for 2(N-1) dependent
variables $a_{1k},a_{2k},k=0,1,...,N-2$. At N=3 such self-dual system
coincides \ in form with the system of sourceless Maxwell equations with
vanishing second invariant
(orthogonal electric and maghetic fields) that provides us with a wide
class of their solutions.

The paper is organized as follows. Basic facts on the Grassmannians are
collected in section 2. Closed differential \ N-1 forms and associated
systems of quasi-linear differential equations are considered in section 3.
\ Solutions of these systems which are Laurent polynomials in one variable
and their relation to N-dimensional integrable equations are studied in
section 4. \ Compact form of these equations and corresponding hierarchies
are discussed in section 5. \ Dual and self-dual quasilinear systems\ are
considered in section 6.
\section{ Grassmannians Gr(N-1,N+1).}
Grassmannian manifold Gr(m,n) is by definition the parameter space for the
totality of m-dimensional linear subspaces $V_{m}$ in the n-dimensional
space $V_{n}$ (see e.g. \cite{HP}, \cite{Sato}, \cite{Tak}). Equivalently
\begin{equation}
Gr(m,n)=\left\{ m\text{-frames in~}V_{n}\right\} /GL(m)
\end{equation}
where an m-frame means an m-tuple $\left\{ p^{0},p^{1},\dots,p^{m-1}\right\}$
of linearly independent vectors with the coordinates $p_{i}^{\beta }$
(i=0,1,\dots,n-1) in a given basis $\left\{ e^{0},e^{1},\dots,e^{n-1}\right\}$
in $V_{n}$. These coordinates can be arranged in the $n\times m$ matrix P
with the elements $p_{i}^{\beta }$. \ The dimension of Gr(m,n) is equal to
m(n-m).

Grassmannain Gr(m,n) can be viewed as an algebraic submanifold of the $\frac{%
n!}{(n-m)!m!}$-dimensional space $\wedge ^{m}V_{n}$ via the correspondence
of a m-frame $\{p^{0}$, $p^{1}$, $\dots$, $p^{m-1}\}$ with the exterior
product $\ p^{0}\wedge p^{1}\wedge\dots\wedge p^{m-1}$ $\in
\wedge ^{m}V_{n}$ \ (the canonical projective embedding). One has
\begin{equation}
p^{0}\wedge p^{1}\wedge \dots \wedge p^{m-1}=\sum_{0\leq
i_{0}\leq \dots \leq i_{m-1}}\pi _{i_{0}\,i_{1}\,\dots\,
i_{m-1}}e^{i_{0}}\wedge e^{i_{1}}\wedge \dots\wedge e^{i_{m-1}}
\label{Pluckercoord}
\end{equation}%
where $\pi _{i_{0}\,i_{1}\,\dots\, i_{m-1}}=\det
(p_{i_{l}}^{k})_{k,l=0,\dots,m-1.}$ These coefficients $\pi _{i_{0}\,i_{1}\,\dots\,
i_{m-1}}$ are called the Pl\"ucker coordinates and satisfy the
Pl\"ucker's relations
\begin{equation}
\sum_{l=0}^{m}(-1)^{l}\pi _{i_{0}\,\dots\, i_{m-2}\,i_{l}}
\pi _{i_{0}^{\prime
}\,\dots\,\check i'_{l}\,\dots\, i_{m}^{^{\prime }}}=0
\label{Plucker}
\end{equation}%
where indices $i_{t}$ and $i_{t}^{^{\prime }}$ range over all possible values
and the notation $\check i'_{l}$ means the deletion of this number. The Pl\"ucker
coordinates completely define embedding of Gr(m,n) into $\wedge ^{m}V_{n}$.

Pl\"ucker's coordinates define also the subspaces $V_{m}$ in $V_{n}$ . \
Namely, a point in $V_{n}$ with the coordinates ($y_{0,}y_{1,}\dots,y_{n-1})$
lies in $V_{m}$ iff (see e.g. \cite{HP}), they obey the system of equations
\begin{equation}
\sum_{l=0}^{m}(-1)^{l}
\pi _{i_{o}\,\dots\, i_{l-1}\,i_{l+1}\,\dots\, i_{m}}y_{i_{l}}=0.
\label{Pluckerspace}
\end{equation}%
In virtue of the Pl\"ucker's relations there are n-m linearly independent
equations among them. \ As a consequence, the system (\ref{Plucker}) is equivalent to
the system of n-m equations
\begin{equation}
y_{\gamma }+\sum_{k=0}^{m-1}a_{\gamma k}y_{k}=0,\qquad \gamma =m,m+1,\dots,n-1.
\label{Plucker01}
\end{equation}%
where $a_{\gamma k}$ are independent affine coordinates in Gr(m,n).

The Grassmannian G(m,n) is projectively equivalent to the Grassmannian
Gr(n-m,n) \cite{HP}. The latter has coordinates $\pi
_{i_{m}\,i_{m+1}\,\dots\,i_{N}}^{\ast }$ defined by the relation $\epsilon
_{i_{0}\,i_{1}\,\dots\,i_{N}}\pi _{i_{m}\,i_{m+1}\,\dots\,i_{N}}^{\ast }=\pi
_{i_{0}\,i_{1}\,\dots\,i_{m-1}}$, where $\epsilon
_{i_{0}\,i_{1}\,\dots\,i_{N}}$ is the totally antisymmetric tensor in N+1
dimensions. The coordinates $\pi _{i_{m}\,i_{m+1}\,\dots\,i_{N}}^{\ast}$ \ obey
equations (\ref{Plucker}) \ and the system dual to (\ref{Pluckerspace}) 
is
\begin{equation}
\sum_{l=0}^{n}\pi _{i_{1}\,\dots\,i_{n-m}\,i_{l}}^{\ast }y_{l}^{\ast }=0
\end{equation}%
This system of equations is equivalent to the following one
\begin{equation}
y_{\gamma }^{\ast }+\sum_{k=0}^{n-m-1}a_{\gamma k}^{\ast }y_{k}^{\ast
}=0,\qquad \gamma =n-m,\dots,n
\end{equation}%
where $a_{\gamma k}^{\ast }$ are independent affine coordinates in Gr(n-m,n).

In this paper we will consider Grassmannians Gr(N-1,N+1) for arbitrary N.
They are the 2(N-1)-dimensional sets of N-1-dimensional linear subspaces in
the N+1-dimensional space defined by codimension two constraint
\begin{equation}
y_{N-1}+\sum_{k=0}^{N-2}a_{1k}y_{k}=0,\quad
y_{N}+\sum_{k=0}^{N-2}a_{2k}y_{k}=0
\label{constraint}
\end{equation}%
with $a_{1k}=(-1)^{k}J^{-1}\pi_{0\,\dots\, k-1\, k+1\, \dots\, N-1}$, 
$a_{2k}=(-1)^{k}J^{-1}\pi_{0\,\dots\, k-1\, k+1\, \dots\, N}$, 
$k=0,\dots,N-2$, $J=\pi _{0\, 1\,\dots\, N-2}$. Affine coordinates 
$a_{1k}$ and $a_{2k}$ are completely defined by the set of N-1 independent
vectors in V$_{N+1}$.

In particular, at N=3 for the Grassmannian Gr(2,4) equations (\ref{Plucker01}) are of
the form
\begin{equation}
y_{2}+a_{11}y_{1}+a_{10}y_{0}=0,\quad y_{3}+a_{21}y_{1}+a_{20}y_{0}=0.
\label{Plin3}
\end{equation}%
Pl\"ucker's coordinates obey the single equation
\begin{equation}
\pi _{01}\pi _{23}-\pi _{02}\pi _{13}+\pi _{03}\pi _{12}=0
\label{P3}
\end{equation}%
and
\[
\pi _{01}=\det \left( 
\begin{array}{cc}
p_{0}^{0} & p_{0}^{1} \\ 
p_{1}^{0} & p_{1}^{1}%
\end{array}%
\right) ,\quad \pi _{02}=\det \left( 
\begin{array}{cc}
p_{0}^{0} & p_{0}^{1} \\ 
p_{2}^{0} & p_{2}^{1}%
\end{array}%
\right) ,\quad \pi _{03}=\det \left( 
\begin{array}{cc}
p_{0}^{0} & p_{0}^{1} \\ 
p_{3}^{0} & p_{3}^{1}%
\end{array}%
\right) , 
\]

\begin{equation}
\pi _{12}=\det \left( 
\begin{array}{cc}
p_{1}^{0} & p_{1}^{1} \\ 
p_{2}^{0} & p_{2}^{1}%
\end{array}%
\right) ,\quad \pi _{13}=\det \left( 
\begin{array}{cc}
p_{1}^{0} & p_{1}^{1} \\ 
p_{3}^{0} & p_{3}^{1}%
\end{array}%
\right) ,\quad \pi _{23}=\det \left( 
\begin{array}{cc}
p_{2}^{0} & p_{2}^{1} \\ 
p_{3}^{0} & p_{3}^{1}%
\end{array}%
\right)
\end{equation}%
where $p^{0}=(p_{0}^{0},p_{1}^{0},p_{2}^{0},p_{3}^{0})$ and $%
p^{1}=(p_{0}^{1},p_{1}^{1},p_{2}^{1},p_{3}^{1})$ are two vectors in $V_{4}$
defining the plane (\ref{constraint}). 
Due to the relations (\ref{constraint}) one has

\begin{equation}
\pi _{02}=-Ja_{11},\pi _{03}=-Ja_{21},\pi _{12}=Ja_{10},\pi
_{13}=Ja_{20},\pi _{23}=J(a_{21}a_{10}-a_{11}a_{20}).
\end{equation}%
The Grassmannian dual to Gr (N-1,N+1) is Gr(2, N+1). It has the same
dimension 2(N-1) as the former one , but  is the set of 2-dimensional
linear subspaces in the N+1-dimensional space V$_{N+1}$ defined by the
constraints
\begin{equation}
y_{\gamma }^{\ast }+a_{\gamma 1}^{\ast }y_{1}^{\ast }+a_{\gamma 0}^{\ast
}y_{0}^{\ast }=0,\qquad \gamma =2,\dots,N.
\label{DualN}
\end{equation}%
At N=3 equations (\ref{Plin3}),(\ref{P3}) 
and their dual equations (\ref{DualN}) have the same form.
\section{Closed differential N-1-forms and systems of
quasi-linear differential equations}
Now we introduce an infinite family of Grassmannians Gr(N-1,N+1)
parametrized by N+1 variables $x_{0},\,x_{1},\,\dots,\,x_{N}$. Thus, all quantities
considered in the previous section become functions of these variables. \ A
geometrical realization of such family is provided, for example, by the
2(N+1)-dimensional cotangent bundle with the local coordinates $%
x_{0},\,x_{1},\,\dots,\,x_{N}$ in the base manifold and cooordinates $%
p_{0},\,p_{1},\,\dots,\,p_{N}$ in cotangent space $T_{N+1}^{\ast}$ and the
Grassmannian Gr(N-1,N+1) in $T_{N+1}^{\ast}$ attached to each point of the
base space.

For each vector $p^{\beta }$ defining the linear N-1 dimensional subspace in 
$V_{N+1}$ there is a naturally associated differential 1-form $\omega
_{\beta }=\sum_{i=0}^{N}p_{i}^{\beta }dx_{i}$ . The projective embedding of
the Grassmannian Gr(N-1,N+1) ( formula (\ref{Pluckercoord})) 
suggests to introduce
differential N-1-form
\begin{equation}
\Omega _{N-1}=\omega _{0}\wedge \dots \wedge \omega _{N-2}
\label{Omegaform0}
\end{equation}%
associated with the member of the family of Grassmannians corresponding to
the point $\left(x_{0},\,x_{1},\,\dots,\,x_{N}\right)$. One has
\begin{equation}
\Omega _{N-1}=\sum_{0\leq i_{0}\leq \dots \leq i_{N-2}}\pi
_{i_{0}\,i_{1}\,\dots\, i_{N-2}}(x)dx_{i_{0}}\wedge dx_{i_{1}}\wedge
\dots \wedge dx_{i_{N-2}}
\label{Pluckerform}
\end{equation}%
where all indices $i_{k}$ take values 0,1,...,N.The homogeneous coordinates $%
\pi _{i_{0}i_{1}\cdot \cdot \cdot i_{N-2}}$ are defined up to the common
factor $\rho $ ,depending on x. So, the form (\ref{Pluckerform}) is defined up to the
scaling transformations $\Omega _{N-1}\rightarrow \rho (x)\Omega _{N-1}$ \
with arbitrary function $\rho (x)$ which can be viewed as the gauge
transformations.

To select a special family of Grassmannians Gr(N-1,N+1) we require that this
form is closed. \ Thus, we require that the Pl\"ucker coordinates obey the
equations
\begin{equation}
\left[ \frac{\partial \pi _{i_{0}\,i_{1}\,\dots\,i_{N-2}}}{\partial x_{i_{N-1}}}%
\right] =0
\label{closedform1}
\end{equation}%
where indices $i_{k}$ take values $0,1,\dots,N$ and 
the bracket $\left[ \,{\dots}\,\right]$
means antisymmetrization over all these indices. This system of $2\frac{%
C_{N-1}^{N+1}}{C_{N-1}^{N}}=N+1$ differential equations toqether with the
algebraic Pl\"ucker relations (\ref{Plucker}) 
form a full system of equations which
characterizes family of Grassmannians Gr(N-1,N+1) for which the form $\Omega
_{N-1}$ is closed.

The system (\ref{closedform1}) 
is equivalent to the system
\begin{equation}
\frac{\partial J}{\partial x_{N-1}}+\sum_{m=0}^{N-2}\frac{\partial (Ja_{1m})%
}{\partial x_{m}}=0,\quad \frac{\partial J}{\partial x_{N}}+\sum_{m=0}^{N-2}%
\frac{\partial (Ja_{2m})}{\partial x_{m}}=0,
\label{gauge}
\end{equation}
\begin{equation}
\frac{\partial a_{1k}}{\partial x_{N}}-\frac{\partial a_{2k}}{\partial
x_{N-1}}+\sum_{l=0}^{N-2}\left( a_{2l}\frac{\partial a_{1k}}{\partial x_{l}}%
-a_{1l}\frac{\partial a_{2k}}{\partial x_{l}}\right) =0,\qquad k=0,1,...,N-2.
\label{subsystem1}
\end{equation}%
The subsystem (\ref{subsystem1}), invariant under the gauge transformations mentioned
above, is the gauge invariant form of the system (\ref{closedform1}). 
Two equations (\ref{gauge})
can be viewed as the equations for the gauge variable J which transforms as $%
J\rightarrow \rho J$ under the gauge transformations. \ It is a
straightforward check \ \ that the equations (\ref{gauge}) are compatible due to the
subsystem (\ref{subsystem1}). 
We note that the system (\ref{gauge},\ref{subsystem1}) can be rewritten in the form
\begin{equation}
D_{1}\ln J+\sum_{m=0}^{N-2}\frac{\partial a_{1m}}{\partial x_{m}}=0,\quad
D_{2}\ln J+\sum_{m=0}^{N-2}\frac{\partial a_{2m}}{\partial x_{m}}=0,
\end{equation}
\begin{equation}
D_{2}a_{1k}-D_{1}a_{2k}=0,\quad k=0,...,N-2
\end{equation}%
where $D_{1}$ and $D_{2}$ are the vector fields
\begin{equation}
D_{1}=\frac{\partial }{\partial x_{N-1}}+\sum_{m=0}^{N-2}a_{1m}\frac{%
\partial }{\partial x_{m}},\quad D_{2}=\frac{\partial }{\partial x_{N}}%
+\sum_{m=0}^{N-2}a_{2m}\frac{\partial }{\partial x_{m}}.
\label{vector_fields}
\end{equation}
So the subsystem (\ref{subsystem1}) 
is equivalent to the commutativity $\left[ D_{1},D_{2}%
\right] =0$ of the vector fields $D_{1}$ and $D_{2}.$

The system (\ref{subsystem1}) of nonlinear equations admits an obvious linearization. \
Indeed, \ for a solution $a_{1k,}a_{2k}$ of this system one can find J
solving linear equations (\ref{gauge}) \ Then for variables \ $J,$ $Ja_{1k,}Ja_{2k}$
one has a linear system which coincides with the system (\ref{closedform1}).

The system (\ref{subsystem1}) 
has another geometrical meaning \ Let us denote by $\mathcal{%
J}$ the family of ideals of the N-1-dimensional linear spaces defined by the
equations (\ref{constraint}) 
and require that $\mathcal{J}$ is the Poisson ideal, i.e.
\begin{equation}
\left\{ \mathcal{J}\text{,}\,\mathcal{J} \right\} \subset \mathcal{J}
\label{Coiso}
\end{equation}%
where $\left\{ ~ \text{,} ~ \right\} $ 
is the standard Poisson bracket with $%
x_{0},x_{1},\dots, x_{N}$;
$y_{0},y_{1},\dots,y_{N\text{ }}$ being the canonical
Darboux coordinates.  Condition (\ref{Coiso}) defines the so-called coisotropic
deformations of the linear algebraic variety defined by equations (\ref{Pluckerspace}) 
\cite{KF},\cite{KO}. Such deformations are described exactly by equations
(\ref{subsystem1}).
Thus, the coisotropic deformations of the ideal $\mathcal{J}$ represent
the necessary and sufficient gauge-invariant conditions for the closedness of
the (N-1)-forms (\ref{Omegaform0}).

At N=2 we have two-dimensional family of straight lines in the 3-dimensional
space,1-form $\Omega _{1}=\sum_{k=0}^{2}p_{k}dx_{k}$ and the conditions of
its closedness are given by equations ($J=p_{0}$)
\[
\frac{\partial p_{0}}{\partial x_{1}}-\frac{\partial p_{1}}{\partial x_{0}}%
=0,\quad \frac{\partial p_{0}}{\partial x_{2}}-\frac{\partial p_{2}}{%
\partial x_{0}}=0,\quad \frac{\partial p_{1}}{\partial x_{2}}-\frac{\partial
p_{2}}{\partial x_{1}}=0 
\]%
or
\bea
&&
\frac{\partial J}{\partial x_{1}}+\frac{\partial (Ja_{10})}{\partial x_{0}}%
=0,\quad \frac{\partial J}{\partial x_{2}}+\frac{\partial (Ja_{20})}{%
\partial x_{0}}=0,
\nn\\
&&
\frac{\partial a_{10}}{\partial x_{2}}-\frac{%
\partial a_{20}}{\partial x_{1}}+a_{20}\frac{\partial a_{10}}{\partial x_{0}}%
-a_{10}\frac{\partial a_{20}}{\partial x_{0}}=0.
\label{subN2}
\eea%
where \ $J=p_{0,}a_{10}=-J^{-1}p_{1},a_{20}=-J^{-1}p_{2}$.

At N=3 one has family of planes (\ref{Plin3}), 2-form $\Omega _{2}$ given by
\begin{eqnarray}
\Omega _{2} &=&J(dx_{0}\wedge dx_{1}-a_{11}dx_{0}\wedge
dx_{2}-a_{21}dx_{0}\wedge dx_{3}+a_{10}dx_{1}\wedge dx_{2}+ \\
&&+a_{20}dx_{1}\wedge dx_{3}-(a_{11}a_{20}-a_{10}a_{21})dx_{2}\wedge dx_{3})
\nonumber
\end{eqnarray}%
and the equations
\begin{equation}
\frac{\partial J}{\partial x_{2}}+\sum_{m=0}^{1}\frac{\partial (Ja_{1m})}{%
\partial x_{m}}=0,\quad \frac{\partial J}{\partial x_{3}}+\sum_{m=0}^{1}%
\frac{\partial (Ja_{2m})}{\partial x_{m}}=0,
\label{subN3a}
\end{equation}
\begin{equation}
\bigskip \frac{\partial a_{1k}}{\partial x_{3}}-\frac{\partial a_{2k}}{%
\partial x_{2}}+\sum_{l=0}^{1}\left( a_{2l}\frac{\partial a_{1k}}{\partial
x_{l}}-a_{1l}\frac{\partial a_{2k}}{\partial x_{l}}\right) =0,\qquad k=0,1.
\label{subN3}
\end{equation}

Finally, we note that at arbitrary N the N-1 form $\Omega _{N-1}$ has a
factorized form
\begin{equation}
\Omega _{N-1}=J\tilde{\Omega }_{N-1}
\label{tildeform}
\end{equation}%
where $\tilde{\Omega }_{N-1}$ is the gauge invariant N-1-form depending
on the affine coordinates $a_{1k},a_{2k}$ only.
\section{Solutions rational in one variable and N-dimensional integrable
systems}
The system (\ref{subsystem1}) of N-1 equations 
for 2(N-1) dependent variables always is
the underdetermined system . In order to make it determined one should
imposes constraints. The simplest constraint is to choose $a_{2k}$ to be
certain functions of $a_{1k},$ i.e. $a_{2k}=\varphi
_{k}(a_{10},\,a_{11},\,\dots,\,a_{1N-2})$, $k=0,1,\dots,N-2.$ In this case the system
(26) becomes the system of N+1-dimensional hydrodynamic type equations
\begin{equation}
\frac{\partial a_{1k}}{\partial x_{N}}+\sum_{m=0}^{N-2}A_{km}\frac{\partial
a_{1m}}{\partial x_{N-1}}+\sum_{m=0}^{N-2}B_{kml}\frac{\partial a_{1m}}{%
\partial x_{l}}=0,\quad k=0,1,\dots,N-2
\end{equation}%
where
\begin{equation}
A_{km}=-\frac{\partial \varphi _{k}}{\partial a_{1m}},\quad B_{kml}=\delta
_{km}\varphi _{l}-a_{1l}\frac{\partial \varphi _{k}}{\partial a_{1m}},\quad
k,m,l=0,1,\dots,N-2.
\end{equation}

At N=2 it is effectively the 1+1-dimensional Hopf-Burgers type equation
\begin{equation}
A\frac{\partial a_{10}}{\partial t}+B\frac{\partial a_{10}}{\partial x_{0}}=0
\end{equation}%
where $A=1-\frac{\partial \varphi _{0}}{\partial a_{10}},B=\varphi
_{0}-a_{10}\frac{\partial \varphi _{0}}{\partial a_{10}}$ and $t=\frac{1}{2}%
(x_{2}-x_{1})$.

Different types of reductions of the system (\ref{subsystem1}) are associated with
the restriction to a special subclass of its solutions, for example, to
solutions which are Laurent polynomials in one variable, say $x_{0}$. \ \
Considering, for example, solutions of the systems (\ref{subsystem1}) of the form
\begin{equation}
a_{1k}=\sum_{n=-m_{1}}^{n_{1}}u_{kn}(x_{1},\dots,x_{N})x_{0}^{n},\quad
a_{2k}=\sum_{n=-m_{2}}^{n_{2}}v_{kn}(x_{1},\dots,x_{N})x_{0}^{n},
\end{equation}%
where $k=0,1,\dots,N-2$,
with appropriate $n_{1},m_{1},n_{2},m_{2}$, one gets the systems of
N-dimensional equations for the functions $u_{kn},v_{kn}$ together with the
corresponding constraints between them. A complete classification of
determined systems which can be obtained in such a way is beyond the scope
of this paper. Here we will present several examples to demonstarte that one
can get number of known integrable systems within this approach.

In the simplest case N=2 with the Ansatz
\begin{equation}
a_{10}=u_{0}+\lambda u_{1},\quad a_{20}=v_{0}+\lambda v_{1}+\lambda ^{2}v_{2}
\end{equation}%
where $\lambda =x_{0},x=(x_{1},x_{2},x_{3})$ \ the system (\ref{subN2}) assumes the
form
\begin{equation}
u_{0x_{2}}-v_{0x_{1}}+u_{1}v_{0}-u_{0}v_{1}=0,\quad
u_{1x_{2}}-v_{1x_{1}}-2u_{0}v_{2}=0,\quad v_{2x_{1}}+u_{1}v_{2}=0.
\end{equation}%
This system implies that \ $u_{1}=-\varphi _{x_{1}}$ where $\varphi =\ln
v_{2}$. Under the constraint $v_{1}=0$ \ the above system becomes
\begin{equation}
\varphi _{x_{1}x_{2}}+2u_{0}e^{\varphi }=0,\quad u_{0x_{2}}-\widetilde{v}%
_{x_{1}}e^{-\varphi }=0
\end{equation}%
where $\widetilde{v}=2v_{0}e^{\varphi }$.  Under the further constraint $%
u_{0}=-\frac{1}{2},\widetilde{v}=2$  one gets the Liouville equation
\begin{equation}
\varphi _{x_{1}x_{2}}=e^{\varphi }.
\end{equation}%
for which
\begin{equation}
a_{10}=-\frac{1}{2}-\lambda \varphi _{x_{1}},\quad a_{20}=e^{-\varphi
}+\lambda ^{2}e^{\varphi }.
\end{equation}

Choosing
\[
a_{10}=-\frac{1}{2}-\lambda \varphi _{x_{1}},\quad a_{20}=e^{-\varphi
}+\lambda ^{2}\varphi _{x_{1}x_{2}}+\lambda ^{3}e^{2\varphi }, 
\]%
one obtains the `higher Liouville' equation \cite{BZM87}
\begin{equation}
\varphi _{x_{1}x_{1}x_{2}}-\varphi _{x_{1}}\varphi _{x_{1}x_{2}}-\frac{3}{2}%
e^{2\varphi }=0
\end{equation}
which in terms of the variable $\Psi =\exp (-\frac{1}{2}\varphi )$ is of the
form
\begin{equation}
\left( \frac{\Psi _{x_{1}x_{1}}}{\Psi }\right) _{x_{2}}-\frac{3}{4}\Psi
^{-4}=0.
\end{equation}

Solutions of the Liouville and higher Liouville equations describe
two-parameter deformations of the straight line defined by the linear system

\begin{equation}
p_{1}+a_{10}p_{0}=0,\quad p_{2}+a_{20}p_{0}=0
\end{equation}%
in the three-dimensional space \ with $a_{10,}a_{20}$ given above. \ In
geometrical terms solutions of these equations describe special classes of
congruences of the lines in the three-dimensional space. In a different
context the interrelation between congruences of lines and system of
equations of hydrodynamic type has been \ studied \ in \cite{agafer},\cite%
{fer}.

In the case N=3 the first nontrivial choice is
\bea
&&
a_{10}=u_{0}(x),\quad a_{11}=u_{1}(x)+\lambda , 
\nn\\
&&
a_{20}=v_{0}(x)+\lambda v_{1}(x),\quad
a_{21}=v_{2}(x)+\lambda v_{3}(x)+\lambda ^{2}.
\label{MSanz}
\eea
Substituting this ansatz into two equations (\ref{subN3}) and denoting 
$x=x_{1}$, $y=x_{2}$, $t=x_{3}$, 
one gets the system of equations
\begin{equation}
u_{0t}-v_{0y}+v_{2}u_{0x}-u_{1}v_{0x}-u_{0}v_{1}=0,
\label{uvv002}
\end{equation}
\begin{equation}
u_{1t}-v_{2y}+v_{0}+v_{2}u_{1x}-u_{1}v_{2x}-u_{0}u_{1}=0,
\label{uvv120}
\end{equation}
\begin{equation}
v_{1}-2u_{0}-v_{2x}-v_{3y}+v_{3}u_{1x}-u_{1}v_{3x}=0,
\label{vu10}
\end{equation}
\begin{equation}
v_{1y}+v_{0x}+u_{1}v_{1x}-v_{3}u_{ox}=0,
\label{vv10}
\end{equation}
\begin{equation}
u_{1x}-v_{3x}=0,
\label{uv13}
\end{equation}
\begin{equation}
u_{0x}-v_{1x}=0.
\label{uv01}
\end{equation}%
Equations (\ref{uv13}), (\ref{uv01}) 
imply that $v_{1}=u_{0,}v_{3}=u_{1}$ and equations
(\ref{vu10}), (\ref{vv10}) become
\begin{equation}
u_{0}+u_{1y}+v_{2x}=0,\quad u_{0y}+v_{0x}=0.
\label{uuv012}
\end{equation}%
Last equation (\ref{uuv012}) implies the existence 
of the function $u$ such that $u_{0}=-u_{x},v_{0}=u_{y}$.
As a result from the first equation (\ref{uuv012}) one
gets $u_{1}=v_{x},v_{2}=u-v_{y}$, where $v$ is some function. Substituting
these expressions into equations (\ref{uvv002}), (\ref{uvv120}), 
one obtaines the system
\begin{equation}
u_{xt}+u_{yy}+u_{x}^{2}+(u-v_{y})u_{xx}+v_{x}u_{xy}=0,
\label{ms1}
\end{equation}
\begin{equation}
v_{xt}+v_{yy}+v_{x}v_{xy}+(u-v_{y})v_{xx}=0.
\label{ms2}
\end{equation}%
It is the Manakov-Santini system found in \cite{MS06}, \cite{MS07}. 
Under the constraint $v=0$ it
reduces into the dKP equation
\begin{equation}
u_{xt}+u_{yy}+(uu_{x})_{x}=0
\end{equation}%
while at $u=0$ it is the equation
\begin{equation}
v_{xt}+v_{yy}+v_{x}v_{xy}-v_{y}v_{xx}=0
\end{equation}%
considered in \cite{MASh02,Pavlov03,MASh04}. 
Thus, the Manakov-Santini system is the
gauge-invariant form of the closedness condition for the 2-form
\[
\Omega _{2}=J(d\lambda \wedge dx-(v_{x}+\lambda )d\lambda \wedge
dy-(u-v_{y}+\lambda v_{x}+\lambda ^{2})d\lambda \wedge dt-u_{x}dx\wedge dy+ 
\]
\begin{equation}
+(u_{y}-\lambda u_{x})dx\wedge dt-(u_{x}u-u_{x}v_{y}+u_{y}v_{x}+\lambda
u_{y})dy\wedge dt)
\end{equation}%
associated with the family of Grassmannians Gr(2,4). \ The function $J$ is a
solution of the system
\begin{equation}
J_{y}-u_{x}J_{\lambda }+(\lambda +v_{x})J_{x}+v_{xx}J=0,
\end{equation}
\begin{equation}
J_{t}+(u_{y}-\lambda u_{x})J_{\lambda }+(u-v_{y}+\lambda v_{x}+\lambda
^{2})J_{x}+(-v_{xy}+\lambda v_{xx})J=0.
\end{equation}

Second example of the 3-dimensional integrable system corresponds to the
choice (again $\lambda =x_{0}$, $x=x_{1}$, $y=x_{2}$, $t=x_{3}$)
\bea
&&
a_{10}=-\lambda (\phi _{y}-\phi _{x}\frac{m_{y}}{m_{x}}),\quad
a_{11}=-\lambda -\frac{m_{y}}{m_{x}},
\nn\\
&&
a_{20}=-e^{\phi }\frac{\phi _{x}}{%
m_{x}},\quad 
a_{21}=-\frac{1}{\lambda }\frac{e^{\phi }}{m_{x}},
\eea%
where $\phi $ and $m$ are functions of $x$, $y$, $t$. The system (\ref{subN3}) 
takes the form
\begin{equation}
(e^{\phi })_{xx}-m_{x}\phi _{ty}+m_{y}\phi _{tx}=0,
\end{equation}
\begin{equation}
e^{\phi }m_{xx}-m_{x}m_{ty}+m_{y}m_{tx}=0.
\end{equation}%
It is the generalization of the 2DTL given in \cite{LVB10Toda}. 
Under the reduction $m=x$
it is the 2DTL equation
\begin{equation}
(e^{\phi })_{xx}-\phi _{ty}=0
\end{equation}%
and at $\phi =0$ it is the equation
\begin{equation}
m_{xx}-m_{x}m_{ty}+m_{y}m_{tx}=0
\end{equation}%
considered in \cite{MASh02}, \cite{MASh04}, \cite{Pavlov03}.

The corresponding 2-form is given by
\[
\Omega _{2}=J(d\lambda \wedge dx+(\lambda -\frac{m_{y}}{m_{x}})d\lambda
\wedge dy+\frac{1}{\lambda }\frac{e^{\phi }}{m_{x}}d\lambda \wedge
dt+\lambda (\phi _{y}-\phi _{x}\frac{m_{y}}{m_{x}})dx\wedge dy+ 
\]
\begin{equation}
+e^{\phi }\frac{\phi _{x}}{m_{x}}dx\wedge dt-\frac{e^{\phi }}{m_{x}}(\lambda
\phi _{x}+\phi _{y})dy\wedge dt)
\end{equation}%
and J is a solution of the system
\[
J_{y}+\lambda (\phi _{y}-\phi _{x}\frac{m_{y}}{m_{x}})J_{\lambda }-(\lambda +%
\frac{m_{y}}{m_{x}})J_{x}+(\phi _{y}-\phi _{x}\frac{m_{y}}{m_{x}}-(\frac{%
m_{y}}{m_{x}})_{x})J=0, 
\]
\begin{equation}
J_{t}-\frac{e^{\phi }\phi _{x}}{m_{x}}J_{\lambda }-\frac{1}{\lambda }\frac{%
e^{\phi }}{m_{x}}J_{x}-\frac{1}{\lambda }(\frac{e^{\phi }}{m_{x}})_{x}J=0.
\end{equation}

Now let us consider the case N=4 and choose the functions $a_{1k}$ and 
$a_{2k}$, $k=0,1,2$, as
\bea
&&
a_{10}=u_{0},\quad a_{11}=u_{1}+\lambda ,\quad a_{12}=u_{2},
\nn\\
&&
a_{20}=v_{0,}\quad a_{21}=v_{1},\quad a_{22}=v_{2}+\lambda
\eea%
where $u_{k}$, $v_{k}$; $k=0,1,2$ are functions of the variables
$x=x_{1}$, $y=x_{2}$, $z=x_{3}$, $t=x_{4}$ and $\lambda =x_{0}$. Substituting this
ansatz into the system (\ref{subsystem1}) with N=4, one gets the equations
\begin{equation}
u_{0t}-v_{0z}+v_{1}u_{0x}+v_{2}u_{0y}-u_{1}v_{0x}-u_{2}v_{0y}=0,
\end{equation}
\begin{equation}
u_{1t}-v_{1z}+v_{0}+v_{1}u_{1x}+v_{2}u_{1y}-u_{1}v_{1x}-u_{2}v_{1y}=0,
\end{equation}
\begin{equation}
u_{2t}-v_{2z}-u_{0}+v_{1}u_{2x}+v_{2}u_{2y}-u_{1}v_{2x}-u_{2}v_{2y}=0,
\end{equation}
\begin{equation}
u_{ky}-v_{kx}=0,\quad k=0,1,2.
\end{equation}%
The last equations imply that
\begin{equation}
u_{0}=u_{x},v_{0}=u_{y},\quad u_{1}=v_{x},v_{1}=v_{y},\quad
u_{2}=w_{x},v_{2}=w_{y}
\end{equation}%
where $w$, $u$, $v$ are functions obeying the equations
\begin{equation}
u_{xt}-u_{yz}+v_{y}u_{xx}+(w_{y}-v_{x})u_{xy}-w_{x}u_{yy}=0,
\label{u}
\end{equation}
\begin{equation}
v_{xt}-v_{yz}+u_{y}+v_{y}v_{xx}+(w_{y}-v_{x})v_{xy}-w_{x}v_{yy}=0,
\label{v}
\end{equation}
\begin{equation}
w_{xt}-w_{yz}-u_{x}+v_{y}w_{xx}+(w_{y}-v_{x})w_{xy}-w_{x}w_{yy}=0.
\label{w}
\end{equation}

The 3-form $\Omega _{3}$ is the sum of 10 terms
\begin{equation}
\Omega _{3}=J(d\lambda \wedge dx\wedge dy-w_{x}d\lambda \wedge dx\wedge
dz-(\lambda +w_{y})d\lambda \wedge dx\wedge dt+(\lambda +v_{x})d\lambda
\wedge dy\wedge dz+\dots)
\end{equation}%
and equations for $J$ are
\begin{equation}
J_{z}+u_{x}J_{\lambda }+(\lambda +v_{x})J_{x}+w_{x}J_{y}+(v_{x}+w_{y})_{x}=0,
\end{equation}
\begin{equation}
J_{t}+u_{y}J_{\lambda }+v_{y}J_{x}+(\lambda +w_{y})J_{y}+(v_{x}+w_{y})_{y}=0.
\end{equation}

The system (\ref{u}-\ref{w}) admits several reductions. 
An obvious one is $u=0$ for which
it is the system of two equations (\ref{v}, \ref{w}) with $u=0$. Less trivial reduction
corresponds to the constraint $v_{x}+w_{y}=0$.  It implies that $v=\Theta
_{y}$ and $w=-\Theta _{x}$. and the system (\ref{u}-\ref{w}) takes the form
\begin{equation}
u_{xt}-u_{yz}+\Theta _{yy}u_{xx}-2\Theta _{xy}u_{xy}+\Theta _{xx}u_{yy}=0,
\label{dun1}
\end{equation}
\begin{equation}
u=\Theta _{zy}-\Theta _{tx}+\Theta _{xy}^{2}-\Theta _{xx}\Theta _{yy}.
\label{dun2}
\end{equation}%
It is the Dunajski system proposed in \cite{Dun02}. 
Under the additional constraint $%
u=0$ one has the Plebanski second heavenly equation \cite{Pleb75}.
\begin{equation}
\Theta _{zy}-\Theta _{tx}+\Theta _{xy}^{2}-\Theta _{xx}\Theta _{yy}=0.
\end{equation}

Integrable systems for any N can be constructed in a similar way. An
interesting problem of finding the systems which describe solutions of the
system (\ref{subsystem1})  
which have poles in one variable, similar to the Calogero-Moser
system for the rational solutions of the Korteweg - de Vries equation \cite{AMM}, 
will be discussed in a separate paper.
\section{Compact form of the hierarchies}
Considering the functions $a_{1k}$ and $a_{2k}$ which are higher order
polynomials in $x_{0}=\lambda $, one gets higher Manakov-Santini and
Dynajski equations (for simplicity we will not consider here two-component
2DTL case which
requires Laurent polynomials, however, it can be also considered in the similar
framework, see \cite{LVB10Toda,LVB12}).
The forms $\Omega _{N-1}$ provide us also with a compact
forms of these hierarchies. These differential forms have rank N-1.
Hence, the condition $d\Omega _{N-1}=0$ implies the existence  of $N-1$
variables $\Psi ^{0},\Psi ^{1},...,\Psi ^{N-2}$ such that (see e.g. \cite{St,Ram})
\begin{equation}
\Omega _{N-1}=d\Psi ^{0}\wedge d\Psi ^{1}\wedge \dots d\Psi
^{N-2}.
\label{Form}
\end{equation}
This means that the components of the vectors $p^{k}$, $k=0,1,\dots,N-2$ which
define N-1-dimensional linear subspaces in the Grassmannian Gr(N-1,N+1)
can be taken as the derivatives $p_{m}^{k}=\frac{\partial \Psi ^{k}}{\partial x_{m}}$,
$k=0,1,\dots,N-2$; $m=0,1,\dots,N$ and equations (\ref{constraint}) take the form 
\begin{equation}
D_{1}\Psi ^{k}=0,\quad D_{2}\Psi ^{k}=0,\quad k=0,1,\dots,N-2
\end{equation}%
where the operators $D_{1}$ and  $D_{2}$ are given by (\ref{vector_fields}).

Since  $\Omega _{N-1}$ is a complicated function in $x_{0}=\lambda $ the
variables $\Psi ^{k}$ are, in general, certain Laurent series in $\lambda $.
An important property of this form is that the N-1-form $\tilde{\Omega 
}_{N-1}$ defined in (\ref{tildeform}) \ is a polynomial function \ in $\lambda $ for
polynomial  $a_{1k}$ and $a_{2k}$. \ Hence, 
\begin{equation}
\left( J^{-1}d\Psi ^{0}\wedge d\Psi ^{1}\wedge \dots d\Psi
^{N-2}\right) _{-}=0
\label{Gen}
\end{equation}%
where $\left( {\cdots}\right) _{-}$ denotes the projection on the part of 
$\left({\cdots}\right)$ 
with negative powers in $\lambda$ and
$$
J=\pi _{0\,1\,\dots\,N-2}=\det
(\p_{l}\Psi^{m})_{m,l=0,\dots,N-2}.
$$
This is the compact
form of the hierarchies of multidimensional integrable systems  considered
in \cite{LVB09}.

Generating relation (\ref{Gen}) implies Lax-Sato equations which
define the evolution of the series $\Psi^{1},\dots,\Psi^{N-2}$ with 
the coefficients depending on $x_m$ ($1\leq m\leq N-2$) with respect to the
times $t_1=x_{N-1}$, $t_2=x_{N}$ (and also higher times corresponding to
commuting flows of the hierarchy). Indeed, the coefficients of the form
(\ref{Form}) can be written as
$\pi _{i_{0}\,i_{1}\,\dots\,i_{N-2}}=\det
(\p_{i_l}\Psi^{m})_{m,l=0,\dots,N-2}$ (compare with relation (\ref{Pluckercoord})).
The functions $\Psi_k$ satisfy identically the relations corresponding
to formula (\ref{Pluckerspace}),  
\begin{equation}
\sum_{l=0}^{N-1}(-1)^{l}
\pi _{i_{0}\,\dots\, i_{l-1}\,i_{l+1}\,\dots\, i_{N}}\p_{i_{l}}\Psi_k=0.
\label{Pluckerspace1}
\end{equation}
An important step is to observe that due to generating relation (\ref{Gen}) the affine
coefficients of linear relations (\ref{Pluckerspace1}) are analytic,
so we obtain nontrivial relations
\begin{equation}
\sum_{l=0}^{m}(-1)^{l}(J^{-1}\pi _{i_{0}\dots i_{l-1}i_{l+1}
\dots i_{m}})_+\p_{i_{l}}\Psi_k=0.
\label{Pluckerspace2}
\end{equation}
Taking the basic relations corresponding to constraints (\ref{constraint}),
we obtain Lax-Sato equations
\bea
&&
\p_{N-1}\Psi^q + \sum_{k=0}^{N-2}a_{1k}\p_{k}\Psi^q=0,
\label{LS1}
\\
&&
\p_{N}\Psi^q +\sum_{k=0}^{N-2}a_{2k}\p_{k}\Psi^q=0
\label{LS2}
\eea%
with 
\beaa
&&
a_{1k}=(-1)^{k}(J^{-1}\pi_{0\,\dots\,k-1\,k+1\,\dots\,N-2\,N-1})_+, 
\\
&&
a_{2k}=(-1)^{k}(J^{-1}\pi_{0\,\dots\,k-1\,k+1\,\dots\,N-2\,N})_+, 
\eeaa 
where $k=0,\dots,N-2$.
Under reasonable
assumptions the coefficients  
$a_{1k}$ and $a_{2k}$ contain only finite number of the coefficients of the series
$\Psi^q$, and the equations (\ref{LS1}), (\ref{LS2}) define the evolution of
these series with respect to times $t_1=x_{N-1}$, $t_2=x_{N}$
(similar equations can be written for the hierarchy of commuting flows).
The compatibility conditions for equations (\ref{LS1}), (\ref{LS2}) considered
as linear equations for $\Psi^q$ are of the form (\ref{subsystem1}), 
and they define
N-dimensional integrable systems for the coefficients of polynomials
$a_{1k}$, $a_{2k}$. They can be also obtained
directly from the condition of the closedness of the form $\tilde\Omega$.

The general multidimensional integrable hierarchy arising
from generating relation of the form (\ref{Gen}) was considered in \cite{LVB09}.
Here we give a brief outline of this construction, emphasizing the connections
with the construction developed in the present work. First, following
\cite{LVB09}, we impose generating relation (\ref{Gen}) on the formal series
of the variable $\lambda$
\bea
&&
\Psi^0=\lambda+\sum_{n=1}^\infty \Psi^0_n(\mathbf{t}^1,\dots,\mathbf{t}^N)\l^{-n},
\label{form0}
\\&&
\Psi^k=\sum_{n=0}^\infty t^k_n (\Psi^0)^{n}+
\sum_{n=1}^\infty \Psi^k_n(\mathbf{t}^1,\dots,\mathbf{t}^N)(\Psi^0)^{-n},
\label{formk}
\eea
where $1\leqslant k\leqslant N-2$, depending on 
$k$ infinite sequences of independent variables
$\mathbf{t}^k=(t^k_0,\dots,t^k_n,\dots)$, $t^k_0=x_k$, $\lambda=x_0$.
Some motivation of introduction of the series of this type and higher
times of the hierarchy is given in \cite{Takasaki89} in the 
simpler case of hyper-K\"ahler hierarchies (vector fields don't contain
a derivative over the spectral variable). The setting of the hierarchy with
infinite number of higher times corresponds to Gr$(N-1,\infty)$.

Slightly modifying notations in the formulae (\ref{LS1}), (\ref{LS2}),
we obtain infinite set of Lax-Sato equations of the hierarchy in the form
\bea
&&
\partial^l_n\Psi^q + (-1)^{N-1}\sum_{k=0}^{N-2}a_n^{lk}\p_{k}\Psi^q=0,
\label{LS00}
\eea
where $l=1,\dots,N-2$, $n=1,\dots,\infty$, 
$\partial^l_n=\frac{\p}{\p t^l_n}$,
\beaa
&&
a_n^{lk}=(-1)^{k} \left(J^{-1}
\left|\frac{D(\Psi^0,\Psi^1,\dots,\Psi^{N-1},\Psi^{N-2})}
{D{({x_0,\dots,\check{x}_{k},\dots,x_{N-2},t^l_n})}}\right|\right)_+,
\eeaa 
and the fraction denotes a Jacobian matrix, $\check x_{k}$ is the absent element. 
Using the series (\ref{form0}),
(\ref{formk}) and estimating the projections, it is possible to simplify these 
expressions and get rid of the derivatives over higher times,
\beaa
&&
(-1)^{N-1}a_n^{lk}=-(-1)^{k}(-1)^l \left(J^{-1}(\Psi^0)^n
\left|\frac{D(\Psi^0,\dots,\check \Psi^{l},\dots,\Psi^{N-2})}
{D{({x_0,\dots,\check x_{k},\dots,x_{N-2}})}}\right|\right)_+.
\eeaa 
The corresponding Lax-Sato equations are
\beaa
&&
\partial^l_n\Psi^q=\sum_{k=0}^{N-2}
(-1)^{k}(-1)^l \left(J^{-1}(\Psi^0)^n
\left|\frac{D(\Psi^0,\dots,\check\Psi^{l},\dots,\Psi^{N-2})}
{D{({x_0,\dots,\check x_{k},\dots,x_{N-2}})}}\right|\right)_+
\p_{k}\Psi^q.
\eeaa
These equations have an evolutionary form and define the dynamics of the series
$\Psi^q$ with the coefficients depending on $x_1,\dots,x_{N-2}$ with respect to
the higher times. The evolutionary form of Lax-Sato equations is connected with 
the special choice of the form of the series (\ref{form0}), (\ref{formk}), it is
a serious argument in favour of this choice.

Using the Jacobian matrix
$$
(\text{Jac}_0)=\left(\frac{D(\Psi^0,\dots,\Psi^{N-2})}
{D{({x_0,\dots,x_{N-2}})}}\right),\quad \det(\text{Jac}_0)=J,
$$
we obtain Lax-Sato equation
in the form introduced in \cite{LVB09},
\bea
&&
\partial^k_n\mathbf{\Psi}=\sum_{i=0}^{N-2}
\left((\text{Jac}_0)^{-1})_{ik} (\Psi^0)^n)\right)_+
{\partial_i}\mathbf{\Psi},\quad
1\leqslant k \leqslant N-2,
\label{genSato}
\eea
where $1\leqslant n\leqslant \infty$,
$\mathbf{\Psi}=(\Psi^0,\dots,\Psi^{N-2})$. It was proved in \cite{LVB09}
that a complete set of Lax-Sato equations (\ref{genSato}) is equivalent to the
generating relation (\ref{Gen}) and that Lax-Sato flows are compatible.
First flows of the hierarchy read
\bea
\partial^k_1\mathbf{\Psi}=(\lambda \partial_k-\sum_{p=1}^{N-2} 
(\partial_k u_p)\partial_p-
(\partial_k u_0)\partial_\lambda)\mathbf{\Psi},\quad 1\leqslant k\leqslant N-2,
\label{genlinear}
\eea
where $u_0=\Psi^0_1$,
$u_k=\Psi^k_1$, $1\leqslant k\leqslant N-2$.
A compatibility condition for any pair of linear equations  
(e.g., with $\partial^k_1$ and $\partial^q_1$, $k\neq q$)
implies closed nonlinear 
N-dimensional  system of PDEs for the set of functions $u_k$, $u_0$,
which can be written in the form
\bea
&&
\partial^k_1\p_q\hat u-\partial^q_1\p_k\hat u+[\p_k \hat u,\p_q \hat u]=
(\p_k u_0)\p_q-(\p_q u_0)\p_k,
\nn\\
&&
\partial^k_1\p_q u_0 - \partial^q_1\p_k u_0 + (\p_k \hat u)\p_q u_0 -
(\p_q \hat u)\p_k u_0=0,
\label{Gensystem}
\eea
where $\hat u$ is a vector field, $\hat u=\sum_{p=1}^{N-2} u_k \p_k$. 
For $N=4$ this system corresponds to the system (\ref{u}-\ref{w}) and,
under a volume-preservation reduction $J=1$, to the Dunajski system
(\ref{dun1}, \ref{dun2}). The case $N=3$ corresponds to the hierarchy connected with
the system (\ref{ms1}, \ref{ms2})
(the Manakov-Santini hierarchy, see \cite{LVB09}).
\section{Dual and self-dual quasi-linear systems}
For the dual Grassmannian Gr(2,N+1) one defines the differential 2-form
\begin{equation}
\Omega _{2}^{\ast }=\omega _{N-1}\wedge \omega _{N}
\end{equation}%
where 1-forms $\omega _{\gamma }=\sum_{i=0}^{N}p_{i}^{\ast \gamma }dx_{i}$.
So,
\begin{equation}
\Omega _{2}^{\ast }=\sum_{i_{0},i_{1}}\pi _{i_{0},i_{1}}^{\ast
}dx_{i_{0}}\wedge dx_{i_{1}}
\end{equation}%
This 2-form \ and N-1-form $\Omega _{N-1}$ are connected by the Hodge star
(duality) operation $\star $
\begin{equation}
\Omega _{2}^{\ast }=\star \Omega _{N-1}
\end{equation}%
where by definition (see e.g. \cite{Ram})
\[
\star (dx_{i_{0}}\wedge dx_{i_{1}}\wedge \dots \wedge
dx_{i_{m}})= 
\]
\begin{equation}
=\frac{1}{(n-m-1)!}\sum_{i_{m+1},...,i_{n}}\epsilon
_{i_{0},i_{1},...,i_{m},i_{m+1},...,i_{n}}dx_{i_{m+1}}\wedge
dx_{i_{m+2}}\wedge \cdot \cdot \cdot \wedge dx_{i_{n}}.
\end{equation}

Operator $\delta $ dual to the exterior differential $d$ is defined as
\begin{equation}
\delta \Omega =(-1)^{mp+m+1}\star d\star \Omega
\end{equation}%
where \ m is the dimension of the space and p is an order of the form $%
\Omega $. \ In particular, $\delta d\star =\star d\delta $ and the operator $%
\Delta =d\delta +\delta d$ is self-dual. \ The differential form $\Omega $
obeying the condition \ $\delta \Omega =0$ ( or $d\Omega ^{\star }=0$ ) is
called co-closed. The form which is closed and co-closed, i.e. $\Delta
\Omega =0$ is refered as the harmonic form (see e.g. \cite{Ram}).

The condition of closedness of the form $\Omega _{2}^{\ast }$ is given by the
system of equations
\begin{equation}
\left[ \frac{\partial \pi _{i_{0},i_{1}}^{\ast }}{\partial x_{i_{2}}}\right]
=0
\label{closedstar}
\end{equation}%
where indices take all values 0,1,...,N. \ In virtue of (\ref{DualN})
this system is
equivalent to
\begin{equation}
\frac{\partial J^{\ast }}{\partial x_{\gamma }}+\frac{\partial (J^{\ast
}a_{\gamma 0}^{\ast })}{\partial x_{0}}+\frac{\partial (J^{\ast }a_{\gamma
1}^{\ast })}{\partial x_{1}}=0,\quad \gamma =2,...,N,
\label{substar}
\end{equation}
\begin{equation}
\frac{\partial a_{jl}^{\ast }}{\partial x_{k}}-\frac{\partial a_{kl}^{\ast }%
}{\partial x_{j}}+\sum_{m=0}^{1}\left( a_{km}^{\ast }\frac{\partial
a_{jl}^{\ast }}{\partial x_{m}}-a_{jm}^{\ast }\frac{\partial a_{kl}^{\ast }}{%
\partial x_{m}}\right) =0,\qquad l=0,1;\quad j,k=2,...,N
\label{substar1}
\end{equation}%
At N=3 the system (\ref{substar}),
(\ref{substar1}) consists from four equations which in the
variables $J^{\ast }$ and $a_{jl}^{\ast }$ have the form similar to the
original system (\ref{subN3a}),(\ref{subN3}).  
In the original variables, due to the relation $%
\pi _{ik}^{\ast }=\sum_{lm=0}^{3}\epsilon _{iklm}\pi _{lm}$, i.e.
\[
\pi _{01}^{\ast }=2\pi _{23}=-2J(a_{21}a_{10}-a_{11}a_{20}),\quad \pi
_{02}^{\ast }=-2\pi _{13}=-2Ja_{20},\quad \pi _{03}^{\ast }=2\pi
_{12}=2Ja_{10}, 
\]
\begin{equation}
\pi _{12}^{\ast }=2\pi _{03}=-2Ja_{21},\quad \pi _{13}^{\ast }=-2\pi
_{02}=2Ja_{11},\quad \pi _{23}^{\ast }=2\pi _{01}=2J
\end{equation}%
it is of the form
\begin{equation}
\frac{\partial J}{\partial x_{1}}-\frac{\partial (Ja_{21})}{\partial x_{3}}-%
\frac{\partial (Ja_{11})}{\partial x_{2}}=0,
\label{st1}
\end{equation}
\begin{equation}
\frac{\partial J}{\partial x_{0}}-\frac{\partial (Ja_{20})}{\partial x_{3}}-%
\frac{\partial (Ja_{10})}{\partial x_{2}}=0,
\label{st2}
\end{equation}
\begin{equation}
\frac{\partial (Ja_{11})}{\partial x_{0}}-\frac{\partial (Ja_{10})}{\partial
x_{1}}+\frac{\partial }{\partial x_{3}}[J(a_{21}a_{10}-a_{11}a_{20})]=0,
\label{st3}
\end{equation}
\begin{equation}
\frac{\partial (Ja_{21})}{\partial x_{0}}-\frac{\partial (Ja_{20})}{\partial
x_{1}}-\frac{\partial }{\partial x_{2}}[J(a_{21}a_{10}-a_{11}a_{20})]=0.
\label{st4}
\end{equation}
Due to the equations (\ref{st1}, \ref{st2}) the equations  
(\ref{st3}, \ref{st4})
are equivalent to the
following
\begin{equation}
\frac{\partial a_{10}}{\partial x_{1}}-\frac{\partial a_{11}}{\partial x_{0}}%
+a_{10}\frac{\partial a_{11}}{\partial x_{2}}+a_{20}\frac{\partial a_{11}}{%
\partial x_{3}}-a_{11}\frac{\partial a_{10}}{\partial x_{2}}-a_{21}\frac{%
\partial a_{10}}{\partial x_{3}}=0,
\label{star3a}
\end{equation}
\begin{equation}
\frac{\partial a_{20}}{\partial x_{1}}-\frac{\partial a_{21}}{\partial x_{0}}%
+a_{10}\frac{\partial a_{21}}{\partial x_{2}}+a_{20}\frac{\partial a_{21}}{%
\partial x_{3}}-a_{11}\frac{\partial a_{20}}{\partial x_{2}}-a_{21}\frac{%
\partial a_{20}}{\partial x_{3}}=0.
\label{star3b}
\end{equation}%
This dual system \ is quite similar to the original system (\ref{subN3}).

If one requires that the form $\Omega _{2}$ is a harmonic one, then one has
system of four equations (\ref{subN3}), (\ref{star3a}),
(\ref{star3b}) for four dependent variables $%
a_{10},a_{11},a_{20},a_{21}$. \ This system has a simple physical meaning.
Indeed, the conditions of closedness and co-closedness of the form $\Omega
_{2} $, i.e. the equations (\ref{closedform}) and (\ref{closedstar})  
are equivalent to
\begin{equation}
\frac{\partial \pi _{ik}}{\partial x_{l}}+\frac{\partial \pi _{kl}}{\partial
x_{i}}+\frac{\partial \pi _{li}}{\partial x_{k}}=0,\quad \ \sum_{k=0}^{3}%
\frac{\partial \pi _{ik}}{\partial x_{k}}=0,\quad i,k,l=0,1,2,3.
\end{equation}%
This system \ can be viewed as the second and first pairs of the sourceless
Maxwell equations for the electromagnetic field tensor $\pi _{ik}$ , i.e. $%
\pi _{0\alpha }=E_{\alpha },\pi _{\alpha \beta }=-\sum_{\gamma
=1}^{3}\epsilon _{\alpha \beta \gamma }H_{\gamma },\alpha ,\beta =1,2,3$
where $\epsilon _{\alpha \beta \gamma }$ is totally antisymmetric
three-dimensional tensor. The Pl\"ucker's relation (18) \ in this terms is \
of the form
\begin{equation}
\sum_{\alpha =1}^{3}E_{\alpha }H_{\alpha }=0.
\end{equation}
Thus, for the harmonic two-form $\Omega _{2}$ the equations for the
Pl\"ucker's coordinates \ coincide with the sourceless Maxwell equations with
perpendicular electric and magnetic fields. This observation provides us
with the variety of solutions of these equations (see e.g.\cite{Landau})
Given such a solution one gets solution of the system 
(\ref{subN3}), (\ref{star3a}), (\ref{star3b}) via
\begin{equation}
a_{10}=-\frac{H_{3}}{E_{1}},\quad a_{11}=-\frac{E_{2}}{E_{1}},\quad a_{20}=%
\frac{H_{2}}{E_{1}},\quad a_{21}=-\frac{E_{3}}{E_{1}}
\end{equation}%
and $J=E_{1}$.

It is a simple check that under the Manakov-Santini Ansatz (\ref{MSanz}) the
system (\ref{subN3}), (\ref{star3a}), (\ref{star3b})  
has only trivial solution.  Rational solutions
of this system will be studied elsewhere.

At N=4 the dual system (\ref{substar}),
(\ref{substar1}) is the system of  9 equations
\begin{equation}
\frac{\partial J^{\ast }}{\partial x_{2}}+\frac{\partial (J^{\ast
}a_{20}^{\ast })}{\partial x_{0}}+\frac{\partial (J^{\ast }a_{21}^{\ast })}{%
\partial x_{1}}=0,
\label{4N1}
\end{equation}
\begin{equation}
\frac{\partial J^{\ast }}{\partial x_{3}}+\frac{\partial (J^{\ast
}a_{30}^{\ast })}{\partial x_{0}}+\frac{\partial (J^{\ast }a_{31}^{\ast })}{%
\partial x_{1}}=0,
\label{4N2}
\end{equation}
\begin{equation}
\frac{\partial J^{\ast }}{\partial x_{4}}+\frac{\partial (J^{\ast
}a_{40}^{\ast })}{\partial x_{0}}+\frac{\partial (J^{\ast }a_{41}^{\ast })}{%
\partial x_{1}}=0,
\label{4N3}
\end{equation}
\begin{equation}
\frac{\partial a_{20}^{\ast }}{\partial x_{3}}-\frac{\partial a_{30}^{\ast }%
}{\partial x_{2}}+a_{30}^{\ast }\frac{\partial a_{20}^{\ast }}{\partial x_{0}%
}-a_{20}^{\ast }\frac{\partial a_{30}^{\ast }}{\partial x_{0}}+a_{31}^{\ast }%
\frac{\partial a_{20}^{\ast }}{\partial x_{1}}-a_{21}^{\ast }\frac{\partial
a_{30}^{\ast }}{\partial x_{1}}=0,
\label{4N4}
\end{equation}
\begin{equation}
\frac{\partial a_{21}^{\ast }}{\partial x_{3}}-\frac{\partial a_{31}^{\ast }%
}{\partial x_{2}}+a_{30}^{\ast }\frac{\partial a_{21}^{\ast }}{\partial x_{0}%
}-a_{20}^{\ast }\frac{\partial a_{31}^{\ast }}{\partial x_{0}}+a_{31}^{\ast }%
\frac{\partial a_{21}^{\ast }}{\partial x_{1}}-a_{21}^{\ast }\frac{\partial
a_{31}^{\ast }}{\partial x_{1}}=0,
\label{4N5}
\end{equation}
\begin{equation}
\frac{\partial a_{20}^{\ast }}{\partial x_{4}}-\frac{\partial a_{40}^{\ast }%
}{\partial x_{2}}+a_{40}^{\ast }\frac{\partial a_{20}^{\ast }}{\partial x_{0}%
}-a_{20}^{\ast }\frac{\partial a_{40}^{\ast }}{\partial x_{0}}+a_{41}^{\ast }%
\frac{\partial a_{20}^{\ast }}{\partial x_{1}}-a_{21}^{\ast }\frac{\partial
a_{40}^{\ast }}{\partial x_{1}}=0,
\label{4N6}
\end{equation}
\begin{equation}
\frac{\partial a_{21}^{\ast }}{\partial x_{4}}-\frac{\partial a_{41}^{\ast }%
}{\partial x_{2}}+a_{40}^{\ast }\frac{\partial a_{21}^{\ast }}{\partial x_{0}%
}-a_{20}^{\ast }\frac{\partial a_{41}^{\ast }}{\partial x_{0}}+a_{41}^{\ast }%
\frac{\partial a_{21}^{\ast }}{\partial x_{1}}-a_{21}^{\ast }\frac{\partial
a_{41}^{\ast }}{\partial x_{1}}=0,
\label{4N7}
\end{equation}
\begin{equation}
\frac{\partial a_{30}^{\ast }}{\partial x_{4}}-\frac{\partial a_{40}^{\ast }%
}{\partial x_{3}}+a_{40}^{\ast }\frac{\partial a_{30}^{\ast }}{\partial x_{0}%
}-a_{30}^{\ast }\frac{\partial a_{40}^{\ast }}{\partial x_{0}}+a_{41}^{\ast }%
\frac{\partial a_{30}^{\ast }}{\partial x_{1}}-a_{31}^{\ast }\frac{\partial
a_{40}^{\ast }}{\partial x_{1}}=0,
\label{4N8}
\end{equation}
\begin{equation}
\frac{\partial a_{31}^{\ast }}{\partial x_{4}}-\frac{\partial a_{41}^{\ast }%
}{\partial x_{3}}+a_{40}^{\ast }\frac{\partial a_{31}^{\ast }}{\partial x_{0}%
}-a_{30}^{\ast }\frac{\partial a_{41}^{\ast }}{\partial x_{0}}+a_{41}^{\ast }%
\frac{\partial a_{31}^{\ast }}{\partial x_{1}}-a_{31}^{\ast }\frac{\partial
a_{41}^{\ast }}{\partial x_{1}}=0.
\label{4N9}
\end{equation}
It is quite different from the original system 
(\ref{gauge}, \ref{subsystem1}) at N=4. 
The
equations (\ref{4N1}, \ref{4N2}, \ref{4N4}, \ref{4N5}) 
form a closed subsystem for the dependent
variables $J^{\ast },a_{20}^{\ast },a_{21}^{\ast },^{\ast }a_{30}^{\ast
},a_{31}^{\ast }$. \ The same is valid for the groups of variables \ $%
J^{\ast },a_{20}^{\ast },a_{21}^{\ast },^{\ast }a_{40}^{\ast },a_{41}^{\ast
} $ and \ \ $J^{\ast },a_{30}^{\ast },a_{31}^{\ast },^{\ast }a_{40}^{\ast
},a_{41}^{\ast }$. So, the whole system (\ref{4N1}-\ref{4N9}) is decomposed into
three independent subsystems. Each of these subsystems coincides with
the system (\ref{subN3a}-\ref{subN3}) for Gr(2,4). 
Such a decomposition remains valid also
for the rational solutions discussed in section 4. So, the simplest rational
solutions of the system (\ref{4N4}-\ref{4N9}) can be constructed as the common
solutions of \ three independent Manakov-Santini or DToda systems. The
subsystem (\ref{4N4}-\ref{4N9}), which is the gauge invariant form of the system
(\ref{closedstar}), describes the coisotropic deformations of the family of the planes
defined by (\ref{DualN}).
It contains twice more equations than the subsystem (\ref{subsystem1}) at
N=4.

Similar situation takes place for \ the dual systems for Gr (N-1,N+1) at N$%
\geq 5$.
\section*{Acknowledgements}
The research of 
LVB was partially supported by the Russian Foundation for Basic Research under grant
no 10-01-00787 and by the President of Russia grant 6170.2012.2 (scientific schools).

\end{document}